\documentstyle[preprint,aps,prl]{revtex}

\def\be{\begin{equation}}
\def\ee{\end{equation}}
\def\bea{\begin{eqnarray}}
\def\eea{\end{eqnarray}}
\def\E{{\bf E}}
\def\B{{\bf B}}
\def\rms{{\rm rms}}
\def\A{{\bf A}}
\def\J{{\bf J}}
\def\k{{\bf k}}
\def\v{{\bf v}}
\def\x{{\bf x}}
\def\L{{\cal L}}
\def\hm{H^{\rm M}}
\def\em{E^{\rm M}}
\def\hmdot{\dot{H}^{\rm M}}
\def\M{{\rm M}}
\def\V{{\rm V}}
\def\ew{{\rm EW}}
\def\eq{{\rm EQ}}

\begin{document}

\draft

\title{Cosmological Magnetic Fields from Primordial Helicity}

\author{George B. Field$^1$ and Sean M. Carroll$^2$  
\\
{\it
$^1$ Harvard-Smithsonian Center for Astrophysics,
60 Garden St., Cambridge, MA  02138, USA
\\
$^2$ Dept. of Physics and Enrico Fermi Institute, University
of Chicago, Chicago, IL 60637, USA 
\\
Email: {\tt gfield@cfa.harvard.edu}, {\tt carroll@theory.uchicago.edu}
}}

\maketitle

\begin{abstract}
Primordial magnetic fields may account for all or part of the fields
observed in galaxies.  We consider the evolution of the magnetic
fields created by pseudoscalar effects in the early universe.  Such
processes can create force-free fields of maximal helicity; we show
that for such a field magnetic energy inverse cascades to larger
scales than it would have solely by flux freezing and cosmic
expansion.  For fields generated at the electroweak
phase transition, we find that the predicted wavelength today can
in principle be as large as $\sim 10$~kpc, and 
the field strength can be as large as $\sim 10^{-10}$~G.

\end{abstract}

\vfill
\eject

\baselineskip=20pt

The origin of galactic and intergalactic magnetic fields is an
unsolved problem \cite{zho}.  The standard $\alpha$-$\Omega$ 
dynamo theory of galactic magnetic fields \cite{rss}
has been criticized for not adequately
taking into account the back reaction of the growing magnetic field,
and in any case, the theory requires a seed field of unknown origin.
There is now an extensive literature examining the possibility that
magnetic fields were created by exotic processes in the early
universe, although most such processes, if they work at all, produce
fields on scales too small to be of interest to astronomy
\cite{torn}.

In magnetohydrodynamics, energy can be transferred from small
to large scales by a process known as the inverse cascade.
As shown in pioneering work by Pouquet and 
collaborators \cite{pfl}, a critical ingredient of the inverse
cascade mechanism is the presence of substantial magnetic
helicity, also known as the Chern-Simons term 
(although non-helical cascades have also been investigated \cite{beos}).
The idea that magnetic helicity may drive an inverse cascade
from microphysical magnetic fields to large-scale cosmological
fields has been advocated by
Cornwall \cite{cornwall}, and investigated subsequently by
Son \cite{son}, who proposed scaling properties we will
verify below.
The present work has two goals:  1.)~To work towards an 
analytic understanding of the cascade process as an initial-value
problem appropriate for cosmology, by studying similarity 
solutions of the MHD equations in the presence of helicity; and
2.)~To apply this understanding to proposed mechanisms which 
create helical primordial fields via pseudoscalar processes in
the early universe.  A preliminary account has been given in
\cite{cf2}.

The early-universe processes we consider can be thought of as
arising from the evolution of an electrically neutral pseudoscalar 
field $\phi$, coupled to electromagnetism through a Lagrange
density 
\be
  \L = -{1\over 4}\phi F_{\mu\nu} \widetilde F^{\mu\nu} = 
  \phi \E\cdot \B\ ,
  \label{lagrangian}
\ee
where $\widetilde F^{\mu\nu} = {1\over 2}\epsilon^{\mu\nu\rho\sigma}
F_{\rho\sigma}$ is the dual electromagnetic field strength
tensor.  (For simplicity we elide the distinction between
electromagnetic and hypercharge fields; they are related by factors
of order unity.)  The pseudoscalar in question may represent an axion
\cite{tw}, a more general pseudo-Goldstone boson \cite{gfc}, or
may model the effect of a chemical potential for right-handed
electron number (in which case the chemical potential $\mu$ is
given by the time derivative of $\phi$).  This last scenario
has been proposed by Joyce and Shaposhnikov \cite{js}, who showed 
that it could lead to fields at the electroweak phase transition
with magnitude $B_\ew\sim 10^{22}$~G and coherent over length
scales $\lambda_\ew\sim 10^{-8}H^{-1}_\ew \sim 2\times 10^{-9}$~cm.

The field equations for electromagnetism in the presence of the 
interaction (\ref{lagrangian}) are (in units with $c=1$)
\bea
  \partial_t \E &=& \nabla\times\B - \J - \dot\phi \B
  - \nabla\phi \times\E \cr
  \nabla\cdot\E &=& \rho_E + \nabla\phi \cdot\B\cr
  \partial_t \B &=& - \nabla\times\E \cr
  \nabla\cdot\B & =& 0\ ,
\eea
along with Ohm's law, $\J = \sigma (\E + \v\times \B)$,
where $\v$ is the velocity of the fluid.  In the cosmological
context of interest here, $\phi$ is effectively homogeneous
($\nabla\phi =0$) and the charge density $\rho_E$ vanishes.
The magnetic field then satisfies
\be
  (\nabla^2 - \partial_t^2)\B = \sigma
  [\partial_t\B - \nabla\times(\v\times\B)] 
  - \dot\phi \nabla\times\B\ .
  \label{eom}
\ee

For the case that $\sigma = 0$ it has been shown that solutions
of (\ref{eom}) can be unstable to exponential growth in the 
magnetic field \cite{cfj}; Garretson, Field and Carroll \cite{gfc} 
considered field production by such a mechanism during inflation,
concluding that the resulting amplitudes were too small to be
of astrophysical interest.
If one ignores the displacement current $\partial_t\E$ and sets
$\dot\phi =0$, (\ref{eom}) reduces to the induction equation of
dissipative MHD.
In the situation of interest here the conductivity is 
non-negligible; the electric field changes slowly
on the timescales of interest, so we neglect $\partial_t\E$;
and the bulk velocity of the fluid is small, so we neglect
$\v$ as well.  Hence
 the appropriate form of (\ref{eom}) is
\be
  (\partial_t - \eta\nabla^2)\B = \eta
  \dot\phi \nabla\times\B\ ,
  \label{eom2}
\ee
where $\eta=1/\sigma$ is the resistivity.

It is useful at this point to go to Fourier space and decompose
$\B(\k)$ into modes of definite helicity (or equivalently, circular
polarization), $\B(\k) = B_+ \hat{u}_+ + B_- \hat{u}_-$;
here $\hat{u}_\pm = \hat{u}_1 \pm i\hat{u}_2$, where $\hat{u}_1$,
$\hat{u}_2$, and $\hat{u}_3 = {\bf k}/k$ form a right-handed
orthonormal basis.
Then solutions to (\ref{eom2}) are of the form
\be
  B_\pm(\k, t) = B_\pm(\k, 0) \exp 
  \left[-\eta\int 
  k(k \mp \dot\phi)\, dt\right]\ ,
\ee
For $\dot\phi$ positive, the $B_+$ modes will
grow exponentially if $k$ is less than $\dot\phi$, with
maximum growth rate for $k = \dot\phi/2$, while the $B_-$
modes decay away.
When $\dot\phi = 0$ the field undergoes Ohmic decay, 
although Joyce and Shaposhnikov
show that this effect is unimportant for the relevant wavelengths
in their scenario \cite{js}.

The expectation value of the magnetic energy density of an
isotropic plasma in a volume $V$ can be expressed in terms of
an energy spectrum $e^\M_k$ as
\be
  \em = {1\over 2}V^{-1} \int_V \B^2\, d^3x 
  = \int_0^\infty e^\M_k \, dk\ ,
\ee
where
\be
  e^\M_k = 2\pi k^2 \langle \B(\k)\cdot \B^*(\k) \rangle\ .
\ee
Similarly, the
expectation value of the magnetic helicity (or 
Chern-Simons) density can be written
\be
  \hm  = V^{-1} \int_V \A\cdot\B\, d^3x 
  = \int_0^\infty h^\M_k \, dk\ ,
\ee
where
\be
  h^\M_k = 4\pi k^2 \langle \A(\k)\cdot \B^*(\k) \rangle\ .
\ee
Note that while $e^\M_k$ is non-negative, $h^\M_k$ can be of
either sign.  The helicity and energy spectra satisfy a realizability
condition,
\be
  |h^\M_k| \leq 2 k^{-1}e^\M_k\ .
  \label{ineq}
\ee
We say the field is ``maximally
helical'' if, for every $k$, $h^\M_k$ is of the same sign and
saturates this inequality.  (See \cite{fplm} for further discussion.)

In Coulomb gauge ($\k\cdot\A = 0$), the modes of the vector
potential and the magnetic field 
are related by $B_\pm = \pm k A_\pm$.  The energy and helicity spectra
are then
\bea
  e^\M_k &=& 4\pi k^2 \left(|B_+|^2 + |B_-|^2\right)\ ,\cr
  h^\M_k &=& 8\pi k \left(|B_+|^2 - |B_-|^2\right)\ .
\eea
Since the $B_+$ modes are amplified and the $B_-$ modes suppressed
by the evolution of $\phi$, only $B_+$ will contribute;
such a field satisfies $h^\M_k = 2 k^{-1}e^\M_k$, and is
therefore maximally helical.

If the spectrum of a maximally helical magnetic field is strongly 
peaked around some wavenumber $k_p$, the configuration will
be substantially force-free: $\J\times\B = 
(\nabla\times\B)\times\B =0$.
This can be seen by considering
\be
  \nabla\times\B(\x) = \int kB_+(\k)\hat{u}_+ e^{i\k\cdot\x}\, d^3k
  \sim k_p \B(\x)\ .
  \label{ffme}
\ee
Force-freedom has been verified in numerical simulations \cite{mfp}.
This condition
plays an important role in the evolution of the fields.
The velocities associated with force-free fields are small, 
protecting them from the Silk damping (photon viscosity) discussed
in \cite{jko} (which implies in addition that they will not
produce detectable distortions in the spectrum of the cosmic
microwave background \cite{jko2}).


We turn now to the principles behind the inverse cascade.
Pouquet {\it et al.} \cite{pfl} studied MHD turbulence in
the eddy-damped quasi-normal Markovian (EDQNM) approximation,
in which eddy damping is used to close the nonlinear equations in 
Fourier space.  These equations preserve the ideal invariants
of total energy and magnetic helicity.  
Pouquet {\it et al.} studied the evolution of the spectra
$e^\M_k$, $h^\M_k$, and $e^\V_k$ (the kinetic energy) under various
conditions.  The presence of magnetic helicity causes a shift of 
magnetic energy from large $k$ to small, as the system attempts
to minimize magnetic energy while conserving magnetic
helicity [equation (\ref{ineq})].  In their study of greatest
interest here, illustrated in Figure 4 of \cite{pfl}, nonhelical
kinetic energy, together with maximally helical magnetic energy, 
are injected at $k=1$ at a constant rate, and $h^\M_k$ is computed 
for $0.016 < k < 16$ at times $0 < t < 200$.  The result is a 
well-defined wave of magnetic energy and helicity that propagates
from $k=1$ downward, reaching $k=0.03$ at $t=200$.  Similar
results have been obtained elsewhere \cite{mfp,bp,brand}.

Here we attempt to quantify this phenomenon analytically so as
to generalize it to the cosmological case, in which magnetic
energy and helicity are injected in a short time.  To do so,
we use equation (3.4) of \cite{pfl}:
\be
  (\partial_t h^\M_k)_{\rm Nonlocal} = k^{-1}\Gamma_k
  \left(h^\V_k- k^2 h^\M_k\right) + \widetilde{\Gamma}_k e^\M_k
  + (\alpha^\V_k - \alpha^\M_k) e^\M_k - 2 \nu^\V_k k^2 h^\M_k\ .
  \label{pfl3.4}
\ee
This equation is obtained by expanding the full EDQNM equations
at a given $k$ to isolate those nonlinear interactions which are
nonlocal in that they involve either $k' < ak$ or $k' > k/a$,
where $a$ is a small parameter.
The first term in (\ref{pfl3.4}), with the transport coefficient in $k$ - space $\Gamma_k$,
represents the so-called Alfv\'en effect, in which
a relatively strong magnetic field at $k' << k$ provides a background on which
excitation at $k$ propagates as an Alfv\'en wave, bringing
$e^\M_k$ into equipartition with $e^\V_k$, and $k^2h^\M_k$ into equipartition 
with the kinetic helicity $h^\V_k$.  The second term is a correction to the 
Alfv\'en effect.  The third term represents the
$\alpha$-effect of mean-field dynamo theory, according to which any inequality
between the kinetic helicity and $k^2 h^\M_k$, the 
current helicity, at $k' >> k$, tends to destabilize $h^\M_k$
and $e^\M_k$ at $k$.  We describe this effect, which is believed to be responsible
for stellar magnetic fields \cite{moffatt}, further below.  The last term, 
turbulent viscosity, can be shown to be negligible in our case because the velocities are small.

The third term is of greatest significance here.  Particularly
important is the appearance of the current helicity
$\langle {\bf b}\cdot\nabla\times {\bf b}\rangle$, whose spectrum is
$k^2 h^\M_k$.  The current helicity was neglected in early
discussions of the $\alpha$-effect, and was first uncovered by Pouquet
{\it et al} \cite{pfl}.  Its importance has been confirmed in \cite{fbc} by
a very different approach, and it plays a crucial role in what follows.
 
To proceed, we identify the dominant term in (\ref{pfl3.4}) by
reference to Figures 4 and 6 of \cite{pfl}.  First, note that at the
peak $k_p$ of the helicity wave at $t=200$, we find 
$e^\V_k\approx 0.05 e^\M_k$, as would occur if the magnetic field is
exerting little force on the fluid.  This is consistent with the
fact that the injected field is maximally helical, hence substantially
force-free according to (\ref{ffme}).  This view is supported by
the fact that in the direct simulation of \cite{mfp}, the 
configurations formed were still $96\%$ maximally helical (hence
substantially force-free) at $t=45$, with $e^\V_k/ e^\M_k \approx 0.02$
at the peak.
It is also consistent with the fact that the field at wavenumbers
less than $k_p$ is very weak, and hence unable to bring about
equipartition of $e^\V_k$ with $e^\M_k$ via the Alfv\'en 
effect.\footnote{The presence of only a very small velocity in
Pouquet's solutions raises the question as to how the field can
evolve in time, absent dissipation.  The authors are currently
investigating this with an approximate model of the Pouquet 
equation.} We conclude that in our case the Alfv\'en effect is weak, and so we neglect the first term in (\ref{pfl3.4}), and with it, the next term, a correction to it. An important implication is that in the helicity
wave, the condition of maximal helicity applies:
\be
  e^\M_k ={1\over 2} kh^\M_k \ .
  \label{maxhelicity}
\ee

We are left with the $\alpha$-term.  In light of (\ref{maxhelicity}),
(\ref{pfl3.4}) can be written 
\be
  (\partial_t h^\M_k)_{\rm Nonlocal} = {1\over 2}
  (\alpha^\V_k - \alpha^\M_k)kh^\M_k \ ,
  \label{3prime}
\ee
demonstrating explicitly the potential of $\alpha^V_k - \alpha^M_k$ to destabilize
$h^\M_k$.  According to equation (3.6) of \cite{pfl},
\be
  \alpha^\V_k - \alpha^\M_k = - {4\over 3} \int_{k/a}^\infty
  \theta_{kqq} (h^\V_q - q^2 h^\M_q)dq\ ,
  \label{4prime}
\ee
where the inverse decay constant $\theta_{kqq}$ is defined below
[equation (\ref{theta})].
{}From Figure 6 of \cite{pfl}, the lack of kinetic energy in the 
helicity wave means that $|\alpha^\V_k| << |\alpha^\M_k|$,
and can be neglected.  Thus
\bea
  (\partial_t h^\M_k)_{\rm Nonlocal} &=& -{1\over2} \alpha^\M_k kh^\M_k \\
  &=& {2\over 3}kh^\M_k \int_{k/a}^\infty \theta_{kqq} 
  q^2 h^\M_q \ dq\ .
  \label{5prime}
\eea
Hence, if $h^\M_k$ is of one sign throughout the wave (as in our
case, where we assume that helicity of only one sign is injected), an instability is possible.
Pouquet {\it et al.} (p.~342) describe how this instability works.
Suppose the wave has reached $k_p = k_1$.
Because $\alpha_k(k_2)$ is due to helicity at $k>k_2$, and
the helicity is stored at $k_1$, wavenumbers $k_2 < k_1$
become unstable, so that $h^\M_k$ and $e^\M_k$ grow there.
The growing  $h^\M_k$ at $k_2$ destabilizes $k_3<k_2$, and so forth.

This can only be a crude explanation, as it relies on the application
of (\ref{pfl3.4}), (\ref{3prime}) and (\ref{4prime}), which are 
based solely on nonlocal interactions. The reader should bear in mind, however, that
the numerical results in \cite{pfl}, represented in Figures 4, 5,
and 6 of that paper, are based on the full EDQNM equations,
not just the nonlocal version in (\ref{pfl3.4}).

The inverse decay constant $\theta_{kqq}$ is given by 
\be
  \theta_{kqq} = (\mu_k + 2\mu_q)^{-1} \ ,
  \label{theta}
\ee
where the eddy-damping rate $\mu_k$ is given by
\bea
  \mu_k  = (\nu+\eta) k^2 &+& 0.26 \left( \int^k_0
  p^2e_p \, dp\right)^{1/2} \nonumber \\
  &+& \sqrt{2\over 3} k\left( \int^k_0 e_p \, dp\right)^{1/2} \ .
  \label{muk}
\eea
The first term represents damping by viscosity $\nu$ and
resistivity $\eta$.  In astrophysical applications $\nu$ 
and $\eta$ are typically of order $10^{-6}$ times the succeeding
terms, so this effect is important only at 
very large wavenumbers;  we will ignore it henceforth.  The second
term parameterizes damping due to self-distortion, and the third
that due to the nonlinear interaction of Alfv\'en waves.

In light of (\ref{maxhelicity}), we can
use (\ref{simil}) to express (\ref{muk}) as
\be
  \mu_k  = (\hm k_p^3)^{1/2} F(\xi)\ ,
\ee
where
\bea
  F(\xi) = & 0.26 & \left(\int^\xi_0\zeta^3 s(\zeta) \, d\zeta
  \right)^{1/2} \cr
  & & +0.82\xi\left(\int^\xi_0\zeta s(\zeta) \, d\zeta\right)^{1/2}\ .
  \label{fofxi}
\eea
We now look for similarity solutions of the form
\be
  h^\M_k(t) = g(t)s(\xi)\ ,
  \label{simil}
\ee
where the shape function $s(\xi)$ depends on the wavenumber
scaled to the peak:
\be
  \xi = {{k}\over{k_p(t)}}\ .
\ee
We normalize the shape function by $\int s(\xi) \, d\xi = 1$,
which allows us to express the time dependence 
of (\ref{simil}) as 
\be
  g(t) = {{\hm(t)}\over{k_p(t)}}\ .
\ee
Given this {\it ansatz} and $\hm(t)$, a solution
will be fully specified by the shape $s(\xi)$ and the peak
wavenumber $k_p(t)$.

Given (\ref{fofxi}), the evolution equation (\ref{5prime})
can be written
\be
  {\hmdot\over{\hm} } 
  -{\dot{k}_p\over k_p} \left(1+\xi{s'\over s}\right)
   = \left({1\over 2} \hm k_p^3\right)^{1/2}G(\xi)\ , 
  \label{evol2}
\ee
where $s' \equiv ds/d\xi$ and
\be
  G(\xi) = {4\over 3} \xi \int^\infty_{\xi/a} 
  {\zeta^2 s(\zeta)\over F(\xi)+2F(\zeta)} \,d\zeta \ .
\ee

Equation (\ref{evol2}) is valid for $k\sim k_p$, in 
particular for $k=k_p(\xi = 1)$, where $s'=0$ by definition.
Thus, $k_p(t)$ solves the differential equation
\be
  {d \over dt} \ln\left({\hm \over k_p}\right)
  = \left({1\over 2} \hm k_p^3\right)^{1/2}G(1)\ , 
  \label{evol3}
\ee
where $G(1)$ is a dimensionless constant of order unity.
Pouquet {\it et al.} \cite{pfl} considered the case
$\hmdot =$~const, so $\hm = \hmdot t$, in which case the
solution to (\ref{evol3}) is
\be
  k_p(t) = k_p(t_i) t_i /t\ ,
  \label{hmdotconst1}
\ee
provided that
\be
  G(1) = \left[{1\over 8} \hmdot k_p^3(t_i) t^3_i \right]^{-1/2}\ .
  \label{hmdotconst2}
\ee
The numerical solution of the exact EDQNM equations
found in \cite{pfl} fits 
(\ref{hmdotconst1})-(\ref{hmdotconst2}) very well.

Reassured that we can find similarity solutions consistent
with the numerical results, we turn to the  $\hmdot = 0$ case,
relevant to cosmology once the pseudoscalar $\phi$ has stopped
evolving.  Then (\ref{evol2}) implies 
that similarity solutions will once again exist, with
\be
  k_p(t) = k_p(t_i) \left({{t_i}\over{t}}\right)^{2/3}\ ,
  \label{kpt}
\ee
provided that
\be
  G(1) = \left[{9\over 8} \hm k_p^3(t_i) t^2_i \right]^{-1/2}\ .
  \label{kpt2}
\ee
Note that
(\ref{kpt}) verifies a result of Son \cite{son}, derived from
a different line of argument.

In addition to the peak wavelength, we also want to know the rms
magnetic field, $B_\rms = \langle\B^2\rangle^{1/2}$.  For our maximally
helical fields we have
\bea
  \langle\B^2\rangle = 
  2\int_0^\infty e^\M_k\, dk 
  &=& \int_0^\infty k h^\M_k\, dk \cr
  &=& \hm k_p(t) \int_0^\infty \xi s(\xi)\, d\xi\ .
\eea
The integral in the last expression is independent of time.  Thus,
when $\hm =$~constant, (\ref{kpt}) implies
\be
  B_\rms(t) = B_\rms(t_i)\left({{t_i}\over{t}}\right)^{1/3}\ .
  \label{brt}
\ee


The preceding discussion has assumed a flat spacetime background;
it is straightforward to adapt these results to an expanding
Robertson-Walker spacetime with metric $ds^2 = -dt^2 + R^2(t)d\x^2
= R^2(t^*)[-dt^{*2} + d\x^2]$, where $R$ is the scale factor
and $t^*$ is the conformal time.  In the radiation-dominated era,
the complete set of MHD equations is conformally invariant (see
for example \cite{sb}), and our results will apply directly to
the conformally transformed quantities
\be
  B_\rms^* = R^2 B_\rms\ ,\quad 
  k_p^* = R k_p\ .
\ee
Thus, the inverse cascade will be characterized by
$k_p^*(t^*)\propto (t^*)^{-2/3}$ and $B_\rms^*(t^*)\propto (t^*)^{-1/3}$,
where the conformal time in the radiation-dominated era is
given by $t^* = 2(t_\eq^{1/2}/R_\eq)t^{1/2}$,
in which the subscript EQ refers to the epoch of matter-radiation 
equality.  The physical quantities at this epoch are therefore
related to their initial values by
\bea
  B_\rms(t_\eq) &=& R_\eq^{-2}  B_\rms^*(t_\eq^*) \cr 
  &=& R_\eq^{-2}\left({{t_i^*}\over{t_\eq^*}}\right)^{1/3} 
  B_\rms^*(t_i^*) \cr
  &=& {{R^2(t_i)}\over{R_\eq^2}} \left({{t_i}\over{t_\eq}}\right)^{1/6} 
  B_\rms(t_i)
\eea
and
\bea
  k_p(t_\eq) &=& R_\eq^{-1} k_p^*(t_\eq^*) \cr
  &=& R_\eq^{-1} \left({{t_i^*}\over{t_\eq^*}}\right)^{2/3} 
  k_p^*(t_i^*)\cr
  &=& {{R(t_i)}\over{R_\eq}} \left({{t_i}\over{t_\eq}}\right)^{1/3} 
  k_p(t_i)\ .
\eea

It is possible that additional inverse cascade occurs in the 
matter-dominated era following $t_{\rm EQ}$. 
Pending further study, we will assume here
that the effect of this is negligible, and so the following estimates of the
characteristic scale of the field today are lower limits.
Thus, for $t>t_{\rm EQ}$ the field is frozen in, the 
wavenumber scales as $R^{-1}$, and the magnetic field as $R^{-2}$.
Their values today are thus
\be
   B_\rms(t_0) = {{R^2(t_i)}\over{R_0^2}} \left({{t_i}\over{t_\eq}}
  \right)^{1/6}  B_\rms(t_i)
\ee
and
\be
  k_p(t_0) = {{R(t_i)}\over{R_0}} \left({{t_i}\over{t_\eq}}\right)^{1/3} 
  k_p(t_i)\ .
\ee
In summary, a maximally helical
primordial magnetic field created at time $t_i$ (during
radiation domination), with initial amplitude $B_\rms(t_i)$ and
initial coherence length $\lambda(t_i) = 2\pi/k_p(t_i)$, will
undergo an inverse cascade, increasing its length scale by
a factor $(t_\eq/t_i)^{1/3}$ over and above stretching due to the
expansion of the universe, while its amplitude is diluted by
an additional factor $(t_i/t_\eq)^{1/6}$.  Note that our estimates
of the present coherence length substantially exceed those of
Son \cite{son}, who terminated the inverse cascade at
$e^+/e^-$ annihilation because of increasing viscosity.  We have
argued that our modes are nearly force-free, with negligible
velocities; viscosity has no effect.

For purposes of illustration, let us consider the fate of a 
magnetic field created at the electroweak phase transition
($T_\ew = 200$~GeV), so $t_i=t_\ew=6\times 10^{-12}$~sec.  
We express the initial coherence length of the
field in terms of the Hubble radius, $\lambda(t_\ew) = f_\lambda
H^{-1}_\ew = 0.4 f_\lambda$~cm, and the initial amplitude in terms
of the total energy density, $B_\rms(t_\ew) = f_B\sqrt{8\pi\rho_\ew}
= 2\times 10^{25}f_B$~Gauss, where $f_\lambda$ and $f_B$ are
dimensionless factors less than unity.  (We have switched here
from Lorentz-Heaviside units to CGS in order to express the magnetic 
field in Gauss.)  The field today will be coherent over scales
\be
  \lambda(t_0) = 5\times 10^{22}f_\lambda{\rm ~cm}
  = 20f_\lambda{\rm ~kpc}\ ,
\ee
with amplitude
\be
  B_\rms(t_0) = 4\times 10^{-10}f_B{\rm ~Gauss} \ .
\ee
In the electroweak case, then, the characteristic length scale of
the field has been amplified by a factor of $(t_\eq/t_\ew)^{1/3}
= 6\times 10^{7}$ more than would be expected for a frozen-in
configuration.  If the initial field is coherent over the Hubble
radius at the electroweak scale ($f_\lambda \sim 1$) and comparable
in energy to the total energy ($f_B \sim 1$), this results
in a length scale of $\sim 20$~kpc and an amplitude of
$\sim 10^{-9}$~Gauss.  According to Dolag {\it et al.} 
\cite{dolag}, the primordial field required to explain Faraday
rotation measures of the Coma cluster is $\sim 10^{-9}$~Gauss.
The observations are consistent with scales of $\sim 60$~kpc.
Thus, if $f_\lambda\sim f_B \sim 1$, primordial helicity could
explain the fields in the Coma cluster.

However, Joyce and Shaposhnikov \cite{js} have estimated that
their scenario for helical field generation at the electroweak
scale results in fields with $f_\lambda \sim 10^{-8}$ and 
$f_B\sim 2\times 10^{-3}$.  While intriguing (for example as a
candidate seed field for a galactic dynamo), these parameters
fall short of providing a sufficient explanation for the fields
seen in galaxies today.  It is therefore worth considering 
variations on this mechanism, perhaps with different dynamics
for the pseudoscalar $\phi$, or field creation at a later epoch
such as the QCD scale.  Mechanisms which create large-amplitude
fields without appreciable helicity will undergo significantly
less (if any) inverse cascade, and thus have a difficult
time leading to fields of astrophysical significance in the 
present universe.

\section*{Acknowledgments}
This work was supported in part by the National Science Foundation
under grant PHY/94-07195 and NASA under grant NAGW-931, and GF
acknowledges discussions at the Institute for Theoretical Physics
Program on Astrophysical Turbulence, April-June, 2000.

\vfill

\end{document}